\documentclass[journal,article,moreauthors]{Definitions/mdpi}

\renewcommand{\linenumbers}{}

\firstpage{1} 
\pubvolume{1}
\issuenum{1}
\articlenumber{0}
\pubyear{2026}
\copyrightyear{2026}

\datereceived{ } 
\daterevised{ } 
\dateaccepted{ } 
\datepublished{ }

\Title{Study of Supernova Neutrinos 
at ESSnuSB\\{\small (ESSnuSB Collaboration)}}

\Author{M. Ghosh $^{1,*}$, R. Mohanta $^{2,*}$ , P. Panda $^{2,*}$ $^{\dagger}$ et al,}

\AuthorNames{J. Aguilar, M.~Anastasopoulos and Firstname Lastname}

\address{%
$^{1}$ \quad Center of Excellence for Advanced Materials and Sensing Devices, Ruđer Bošković Institute, 10000 Zagreb, Croatia; mghosh@irb.hr\\ 
$^{2}$ \quad School of Physics, University of Hyderabad, Hyderabad 500046, India; rmsp@uohyd.ac.in; ppapia93@gmail.com}

\corres{Corresponding authors: mghosh@irb.hr, rmsp@uohyd.ac.in, ppapia93@gmail.com}

\firstnote{Full author list is provided in the Appendix A}  
 
\abstract{In this paper, we have studied the sensitivity of the ESSnuSB far detector to supernova neutrinos. ESSnuSB is a proposed long-baseline neutrino experiment in Sweden, which will use a 538 kt water Cherenkov detector to probe the leptonic phase $\delta_{\rm CP}$ by studying the second oscillation maximum. However, given the very large detector volume, it will have an excellent sensitivity to supernova neutrinos if a supernova explosion occurs during the run-time of ESSnuSB. Motivated by this, we first estimate the expected event rates at the ESSnuSB far detector for three different supernova flux models and then we probe its capability to distinguish these flux models. Additionally, we also investigate the impact of systematic errors and detector efficiency. Our results show that depending on the model of the supernova neutrinos, the expected number of events detected at Earth varies significantly. Our results also show that the ESSnuSB far detector may have excellent potential in distinguishing these flux models depending upon the distance of the supernova explosion, systematic errors and detector efficiency.  }


\begin{document}

\section{Introduction}

In a core-collapse supernova \cite{Burrows:2000mk,Horiuchi:2017qja}, the core of the star collapses and releases a large amount of energy. During this process, neutrinos of different flavors are produced with energies of the order of a few tens of MeV \cite{Bethe:1990mw}. Studying  supernova neutrinos can provide insights into the structure of the star as well as various properties of neutrinos ~\cite{Scholberg:2012id,Giunti:2007ry,Deng:2023twb,Halzen:2009sm, Panda:2024avc}. In this paper, we study the sensitivity to supernova neutrinos at the ESSnuSB far detector (FD) \cite{Alekou:2022emd}. So far, neutrinos have been detected only from the SN1987A supernova, which occurred in Large Magellanic Cloud at a distance of 50 kpc from Earth \cite{DedinNeto:2023hhp,Lagage:1987xu}. Three detectors on Earth: the water Cherenkov detector Kamiokande II \cite{Kamiokande-II:1987idp}, the water Cherenkov detector Irvine-Michigan-Brookhaven (IMB)~\cite{IMB:1987klg} and the scintillator detector Baksan~\cite{Novoseltseva:2009cr}, detected only a total of 24 neutrino events. Due to its high sensitivity, the future experiment ESSnuSB, which will use a 538 kt water Cherenkov detector as its far detector, will be able to detect a large number of neutrino events if a supernova explosion occurs in the future. This will provide an excellent opportunity to understand various phenomena related to neutrino physics.

There are several models to calculate supernova neutrino fluxes, each based on different assumptions. One of the earliest models is the Livermore model \cite{Totani:1997vj}, which uses one-dimensional simulations based on SN1987A data. It provides neutrino fluxes from the beginning of a collapse up to 18 seconds after the core bounce. Another model is the GVKM (Gava-Volpe-Kneller-McLaughlin) model \cite{Gava:2009pj}, which includes collective effects and shock wave effects in neutrino propagation. It uses the $S$-matrix method along with realistic density profiles. A more realistic model is the Garching electron-capture supernova (ECSN) model \cite{Hudepohl:2009tyy}, where an $8.8 M_\odot$ supernova is simulated using a spherically symmetric framework. This model follows the full evolution of the supernova, including the formation of the neutron star.

In this paper, we first estimate the event rates at the ESSnuSB FD for the Livermore, GVKM, and Garching flux models. In these calculations, neutrino flavor oscillations inside the supernova are taken into account. We then study the ability of the ESSnuSB FD to distinguish between these three flux models within the standard scenario. Additionally, we examine the effects of systematic uncertainties, and detector efficiency.

The structure of this paper is as follows. First, the theoretical background of supernova neutrino oscillations is presented. Next, the simulation details for the ESSnuSB FD are described. This is followed by a discussion of the results, and finally, the conclusions are presented.

\section{Theoretical background}

For supernova neutrinos, the primary neutrino spectra can be parameterized in a flavor-dependent way as
\cite{Keil:2002in,Dasgupta:2008my,Scholberg:2012id}
\begin{equation}
    \Phi_{\nu} (E) = \mathcal{N} \left( \frac{E_{\nu}}{\langle E_{\nu} \rangle} \right)^{\alpha} e^{-(\alpha+1) \frac{E}{\langle E  _{\nu} \rangle}}\;,
\end{equation}
where $\alpha$ denotes the pinching parameter, $\langle E_{\nu} \rangle$ represents the average neutrino energy and $\mathcal{N}$ is the normalization constant expressed as,

\begin{equation}
    \mathcal{N}= \frac{(\alpha+1)^{\alpha+1}}{\langle E_{\nu} \rangle \Gamma(\alpha+1)}\;,
    \label{nor}
\end{equation}
with $\Gamma$ denoting the Euler Gamma function. The neutrino flux ($F_{\nu}^0$) at the neutrinosphere for each flavor can be expressed in terms of $\Phi_{\nu}(E)$ as,

\begin{equation}
    F_{\nu}^0 = \frac{L_{\nu}}{\langle E \rangle_{\nu}} \Phi_{\nu} (E)\;,
    \label{flux}
\end{equation}
where $L_{\nu}$ denotes the luminosity of neutrinos of a given flavor. In the standard three-flavor scenario, after accounting for neutrino oscillations inside the core and at the surface of the star, the resulting fluxes can be written as,
 \cite{Dighe:1999bi,Panda:2023rxa}
\begin{eqnarray}
\label{eq-flux}
    &&F_{\nu_e} = p F_{\nu_e}^0 + (1-p) F_{\nu_x}^0, \\
    \label{eq-1}
   && F_{\bar{\nu}_e} = \bar{p} F_{\bar{\nu}_e}^0 + (1- \bar{p} ) F_{\nu_x}^0 ,\\
   \label{eq-2}
&&   2F_{\nu_x} = (1-p)F_{\nu_e}^0 + (1+p) F_{\nu_x}^0, \\
\label{eq-3}
&& 2F_{\bar{\nu}_x} = (1-\bar{p})F_{\bar{\nu}_e}^0 + (1+\bar{p}) F_{\bar{\nu}_x}^0\;,
\end{eqnarray}
where $p$ and $\bar{p}$ represent the survival probabilities of $\nu_e$ and $\bar{\nu}_e$, respectively. The expressions for these survival probabilities, for both mass hierarchies, are given in Table~\ref{survival} \cite{Dighe:1999bi}.
 In normal hierarchy of the neutrinos, we have $m_1 < m_2 \ll m_3$, where $m_1$, $m_2$ and $m_3$ being the masses of the neutrinos. The inverted hierarchy of the neutrino masses corresponds to the case when $m_3 \ll m_1 < m_2$. From the table, it can be seen that the neutrino oscillation probabilities relevant for supernova neutrinos depend on the mixing angles $\theta_{13}$ and $\theta_{12}$. In this study, we assume that there are no significant oscillation effects on the neutrino flux during propagation from the supernova to Earth. We also neglect any Earth matter effects on the neutrino flux (see \cite{Seadrow:2018ftp} for more details).
 
\begin{table}[htbp]
   \centering
  \scalebox{1.0}{
    \begin{tabular}{|c|c|c|}
    \hline
        Hierarchy & $p$ & $\bar{p}$  \\
        \hline
        Normal & $\sin^2 \theta_{13}$ & $\cos^2 \theta_{12} \cos^2 \theta_{13}$  \\
        \hline
         Inverted & $\sin^2 \theta_{12} \cos^2 \theta_{13}$  &  $\sin^2 \theta_{13}$\\
        \hline
    \end{tabular}}
    \vspace{0.2cm}
    \caption{Survival probability expressions of neutrino ($p$) and antineutrino ($\bar{p}$) fluxes for two cases: normal hierarchy and inverted hierarchy.}
    \label{survival}
\end{table}

\begin{table}[htb]
    \centering
     \scalebox{1.0}{ \begin{tabular}{|c|c|}
    \hline
        Oscillation  Parameters & Best-fit values \cite{Esteban:2024eli} \\
        \hline
        $\theta_{12}$ & $33.76^{\circ}$\\
        \hline
        $\theta_{13}$  & $8.62^{\circ}$ \\
        \hline
        $\theta_{23}$  & $43.29^{\circ}$ \\
        \hline
        $\Delta m_{31}^2 ~(\rm eV^2)$  & $2.511 \times 10^{-3}$\\
        \hline
        $\Delta m_{21}^2 ~(\rm eV^2)$   & $7.537 \times 10^{-5}$ \\
        \hline
    \end{tabular}}
    \caption{Values of oscillation parameters used in the simulation.}
    \label{table2}
\end{table}

It is important to note that, in addition to the Mikheyev-Smirnov-Wolfenstein (MSW) effect, other processes can also influence the flavor oscillations of supernova neutrinos. One such effect arises from neutrino self-interactions, leading to collective flavor transitions \cite{Giarnetti:2026kcy}. These transitions can be classified as slow \cite{Mirizzi:2015eza,Chakraborty:2016yeg,Horiuchi:2018ofe} or fast \cite{Tamborra:2020cul}, depending on their timescales. The slow collective effect can lead to either a spectral swaps which happens only for inverted ordering at 8 MeV \cite{Chiu:2013dya} or there can be multiple spectral splits at different energies and also for normal ordering \cite{Dasgupta:2009mg,Friedland:2010sc}. It is now believed that spectral swaps due to slow effects are suppressed in realistic situations \cite{Sarikas:2011am,Chakraborty:2011nf}. Further, several recent studies have found that the fast instabilities lead to flavour equilibrium prior to the MSW region \cite{Bhattacharyya:2022eed,Wu:2021uvt,Richers:2021nbx,Richers:2021xtf,Sigl:2021tmj}.
The study of these collective effects is still an active area of research, and their full impact on neutrino flavor conversion is not yet completely understood. However, detailed multi-angle studies have shown that energy-dependent spectral features tend to be smoothed out when considering time-integrated spectra after the core bounce, resulting in only small corrections \cite{Lunardini:2012ne,Hajjar:2023knk, Lindner:2002wm}. In recent years, several studies \cite{Panda:2023rxa, Panda:2024avc, Gaba:2024asp} have performed analyses without including collective oscillation effects. Similarly, some experimental collaborations, such as DUNE \cite{DUNE:2023rtr} and T2HK \cite{Hyper-Kamiokande:2021frf}, have also carried out supernova neutrino sensitivity studies without taking collective effects into account. Therefore, in this analysis, we neglect the effects of collective neutrino transitions.

\section{Simulation details}

Currently, we do not have a realistic detector simulation of the ESSnuSB FD for supernova neutrinos. The goal of this paper is to consider a simplistic assumption of the detector properties and study the phenomenology of supernova neutrinos. For the analysis of supernova neutrinos, we use the SNOwGLoBES (Supernova Neutrino Observatories with GLoBES) \cite{github} software. This tool is based on the GLoBES (General Long Baseline Experiment Simulator) package \cite{Huber:2004ka, Huber:2007ji} and is specifically designed for supernova neutrino studies. SNOwGLoBES calculates event rates using inputs as neutrino fluxes, interaction cross sections, and detector response functions.

For ESSnuSB FD, we consider a 538 kt water Cherenkov detector. We further assume a signal efficiency of 95\% and a Gaussian energy resolution of 15\% unless otherwise mentioned. In this analysis, we consider supernova neutrinos in the energy range from 0.5 MeV to 100 MeV, using 200 true energy bins and 200 sampling bins. We do not include any background in our study. This is because, for a Galactic supernova burst, background rates in current and future experiments are expected to be very low. Possible background sources include radioactivity, cosmic rays, $\bar{\nu}_e$, and solar $\nu_e$. Additional contributions may also come from low-energy atmospheric neutrinos and antineutrinos. However, most of these backgrounds can be effectively reduced by placing the detector underground. For a Galactic supernova, approximately $10^5$ neutrino events are expected to be detected within a short time interval of a few tens of seconds. Compared to such a large burst of signal events, the contribution from the above-mentioned backgrounds is expected to be negligible. More details about supernova neutrino backgrounds can be found in Ref.~\cite{Scholberg:2012id}.

Having ultra-pure water as the detector material, the main channel for supernova neutrino detection is inverse beta decay (IBD). In IBD, an incoming $\bar{\nu}_e$ interacts with a proton of the water molecule to form a neutron and a positron:
\begin{equation}
    \bar{\nu}_e + p \rightarrow n + e^+.
 \end{equation}
In ESSnuSB detector, as almost $90 \%$ of events are IBD, we therefore consider only IBD process in our analysis. 

\section{Results}

\subsection{Discussion on supernova fluxes}

\begin{figure}[h]
    \centering
    \includegraphics[width=0.49\linewidth]{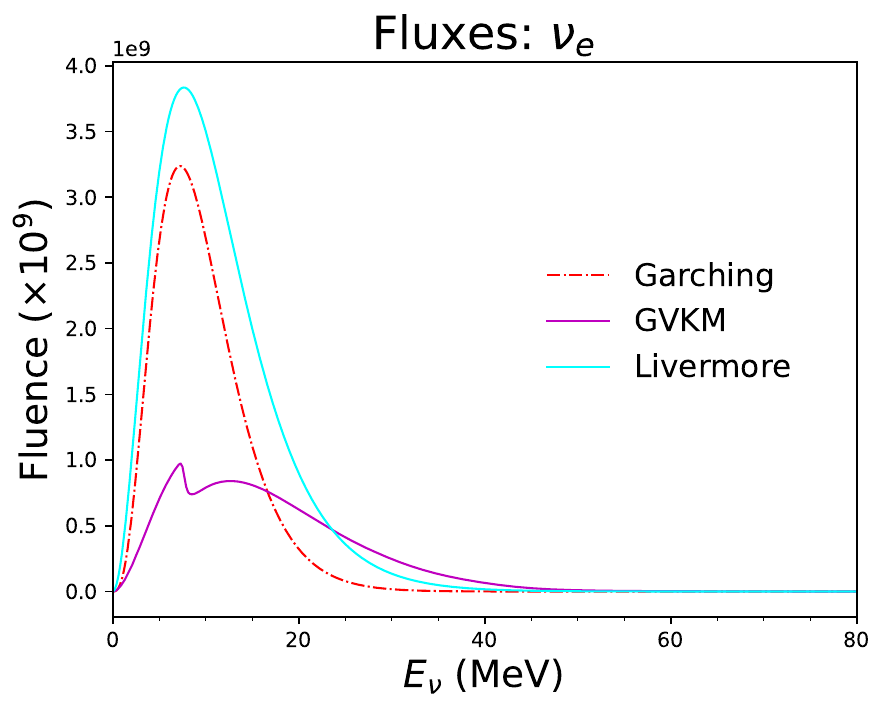}
    \includegraphics[width=0.49\linewidth]{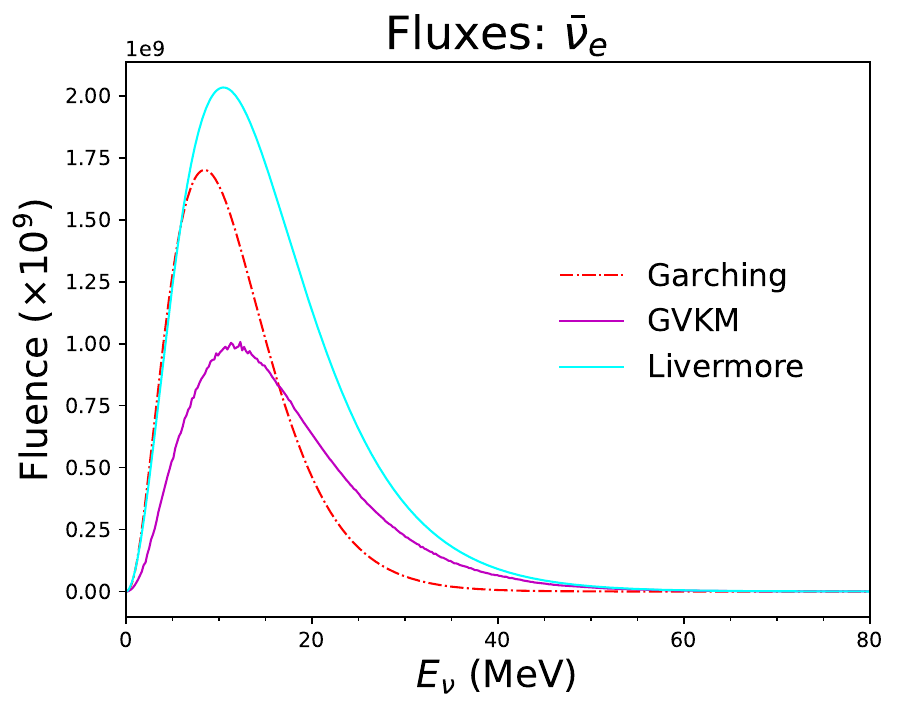} \\
    \includegraphics[width=0.49\linewidth]{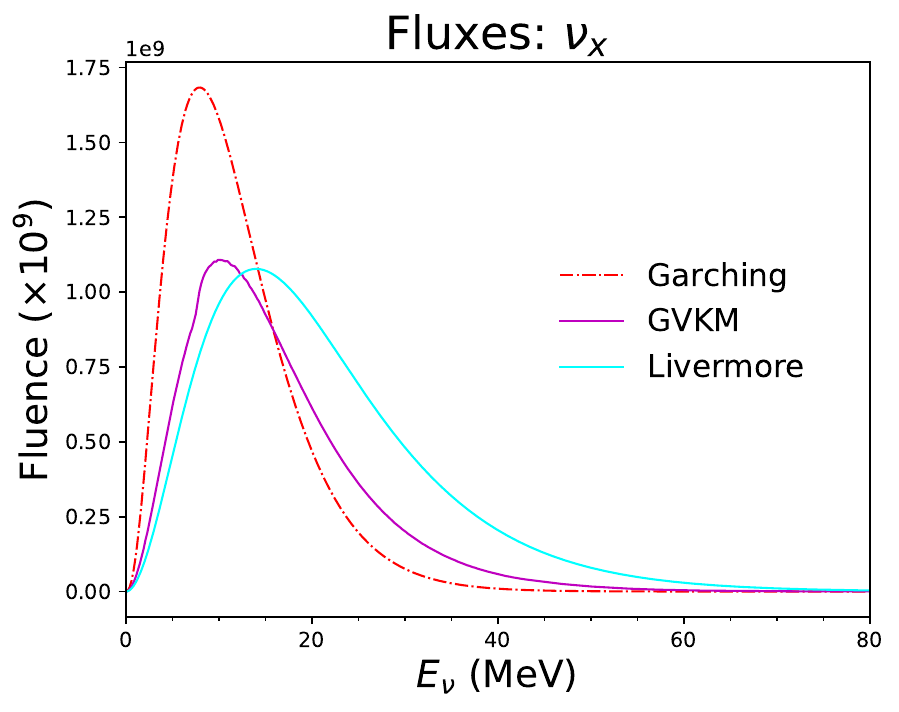}
    \includegraphics[width=0.49\linewidth]{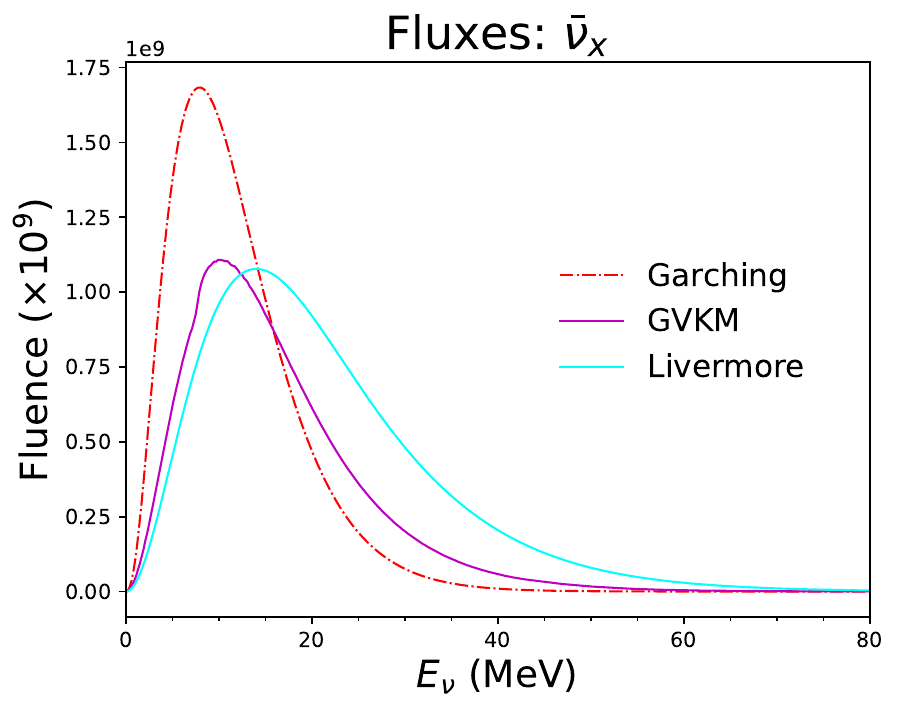}
    \caption{Unoscillated fluence for three different supernova flux models. In each panel, the red curve is for Garching model, the magenta curve is for GVKM model and the cyan curve is for Livermore model.}
    \label{sn_flux}
\end{figure}

In Fig.~\ref{sn_flux}, we show the unoscillated time integrated fluxes, i.e., fluences for a 10 kpc supernova distance for three different supernova flux models as mentioned previously. The top-left panel is for $\nu_e$, the top right is for $\bar{\nu}_e$, the bottom-left is for $\nu_x$ and the bottom-right is for $\bar{\nu}_x$. The color codes used in the figure are indicated in the legend. From the figure we understand that although the shape of the fluxes for different models are quite similar, the overall normalization is significantly different. For $\nu_e$ and $\bar{\nu}_e$, all the three models have different size of the fluxes, whereas for $\nu_x$ and $\bar{\nu}_x$, the predictions of the GVKM and Livermore models are similar but that of  Garching model is very different. Therefore, for supernova neutrinos, the prediction for the number of events will vary significantly depending on the flux models, and thus, it is very important to study the capability of an experiment to distinguish between different flux models. 

\begin{figure}[h]
    \centering
    \includegraphics[width=0.49\linewidth]{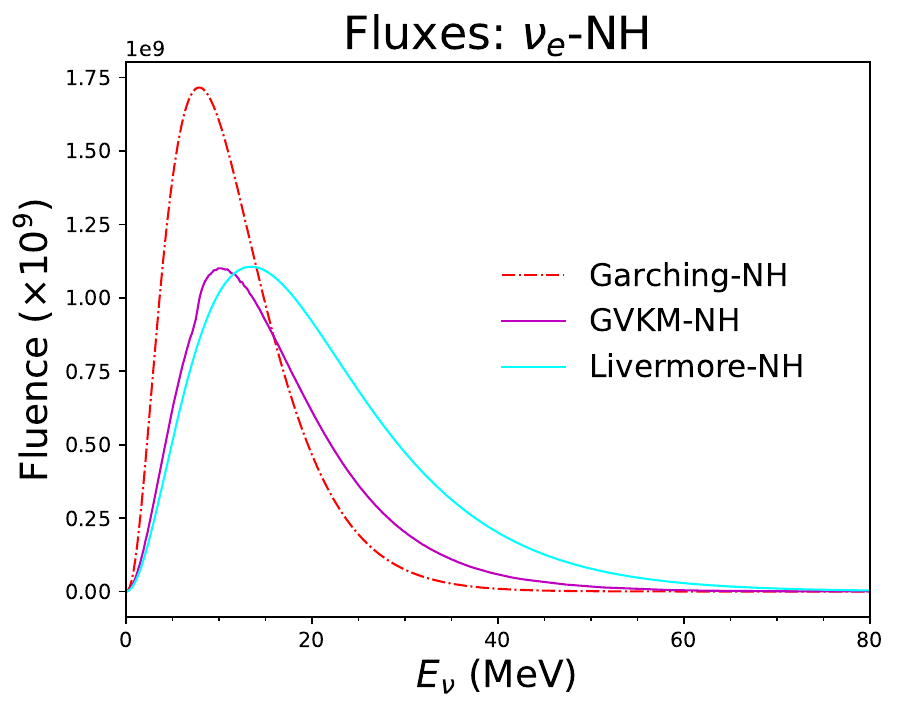}
    \includegraphics[width=0.49\linewidth]{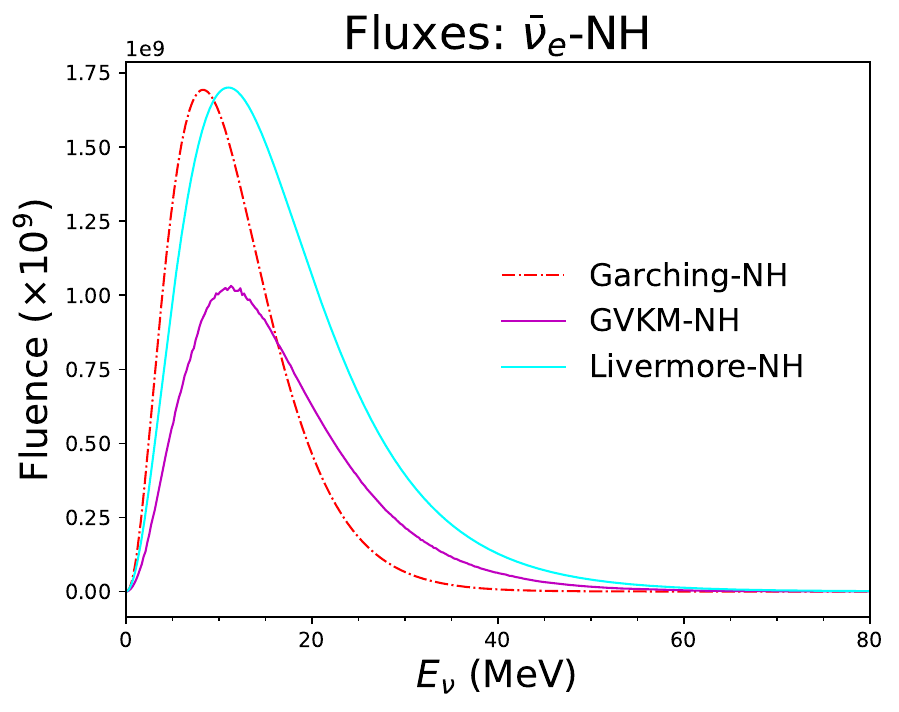} \\
    \includegraphics[width=0.49\linewidth]{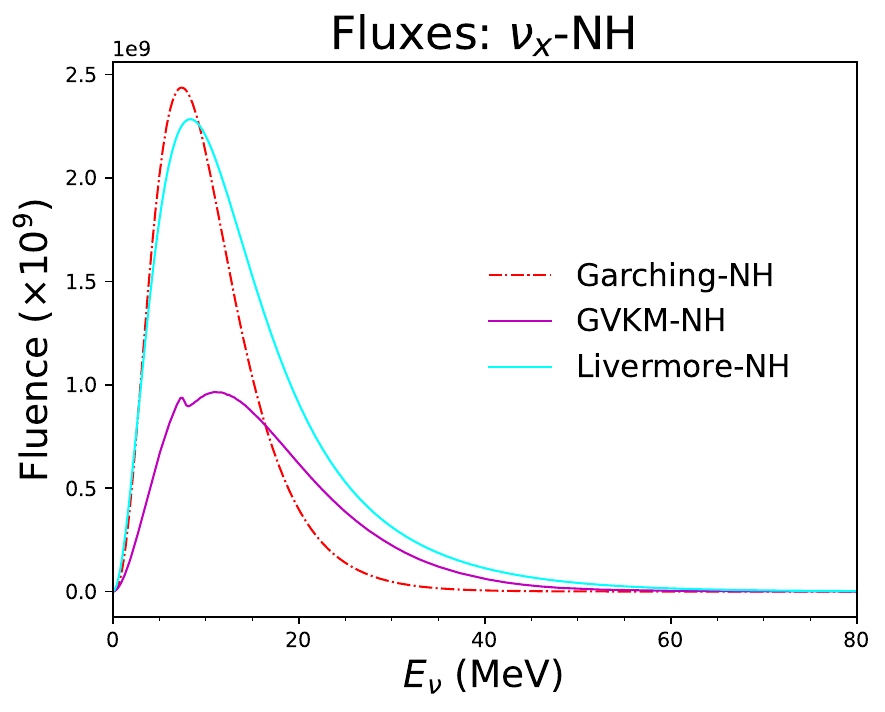}
    \includegraphics[width=0.49\linewidth]{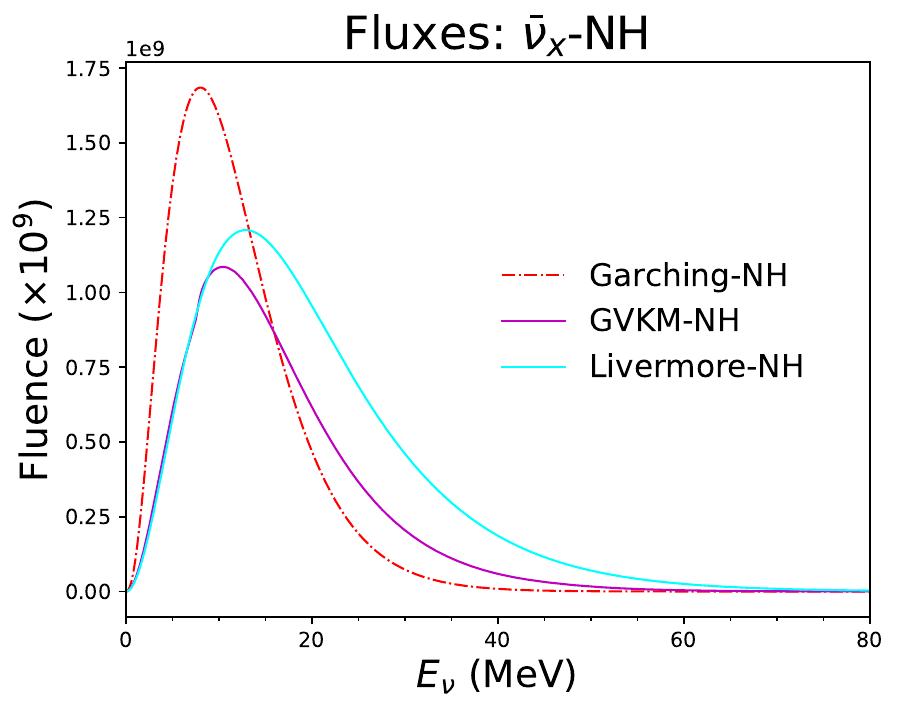}
    \caption{Oscillated fluxes for three different supernova flux models in the normal mass hierarchy of the neutrinos. In each panel the red curve is for the Garching model, the magenta one is for the GVKM model and the cyan  is for the Livermore model.}
    \label{sn_oscillated_flux_nh}
\end{figure}

\begin{figure}[h]
    \centering
    \includegraphics[width=0.49\linewidth]{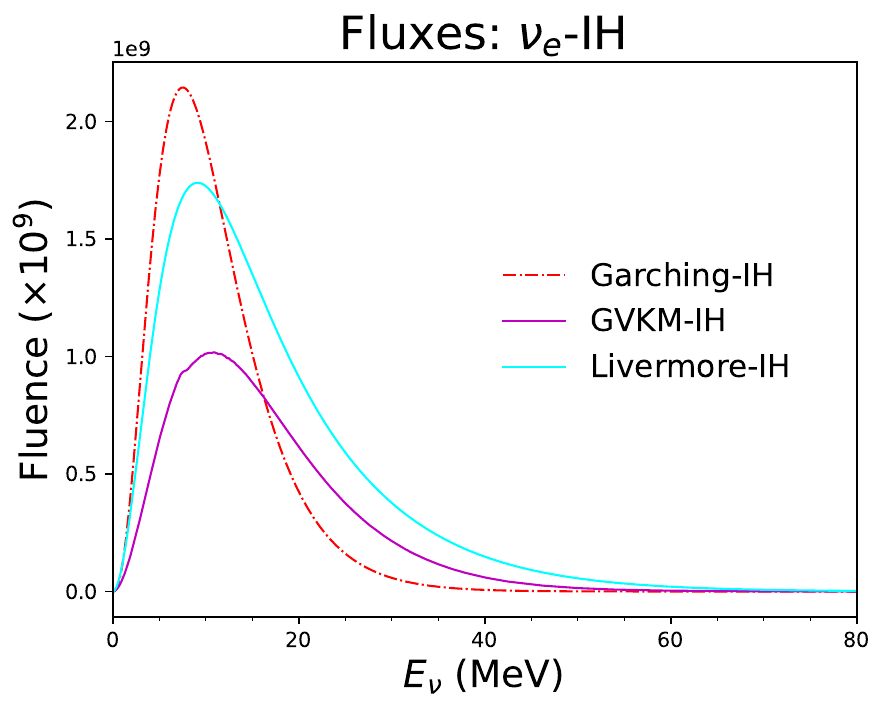}
    \includegraphics[width=0.49\linewidth]{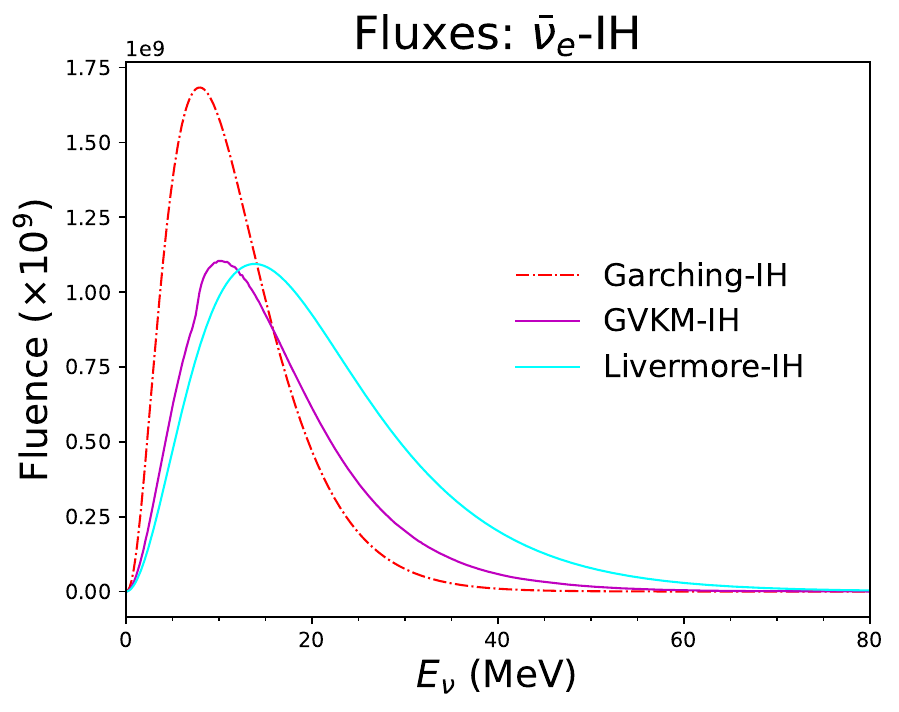} \\
    \includegraphics[width=0.49\linewidth]{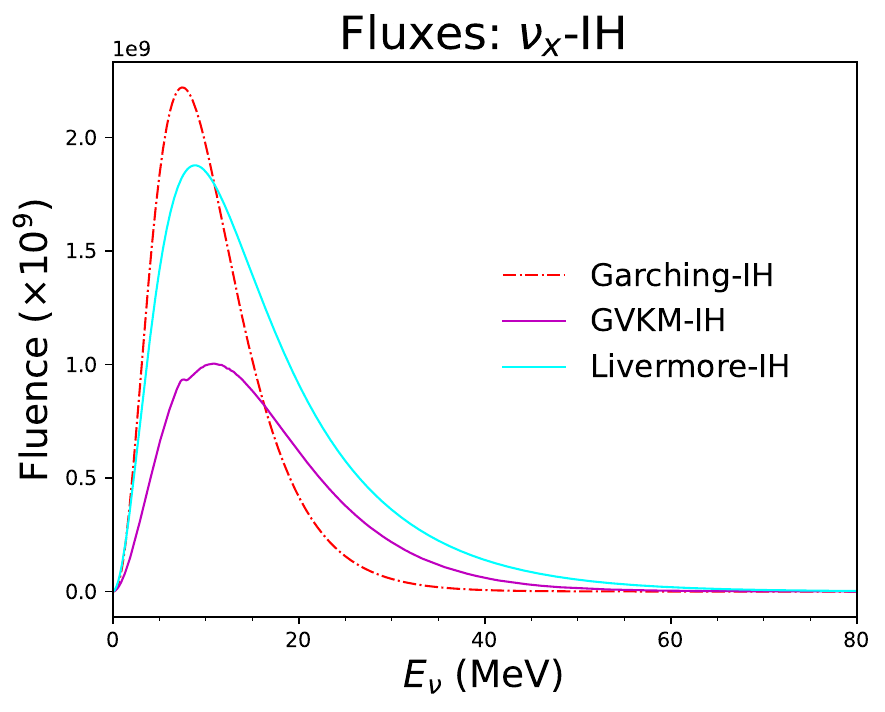}
    \includegraphics[width=0.49\linewidth]{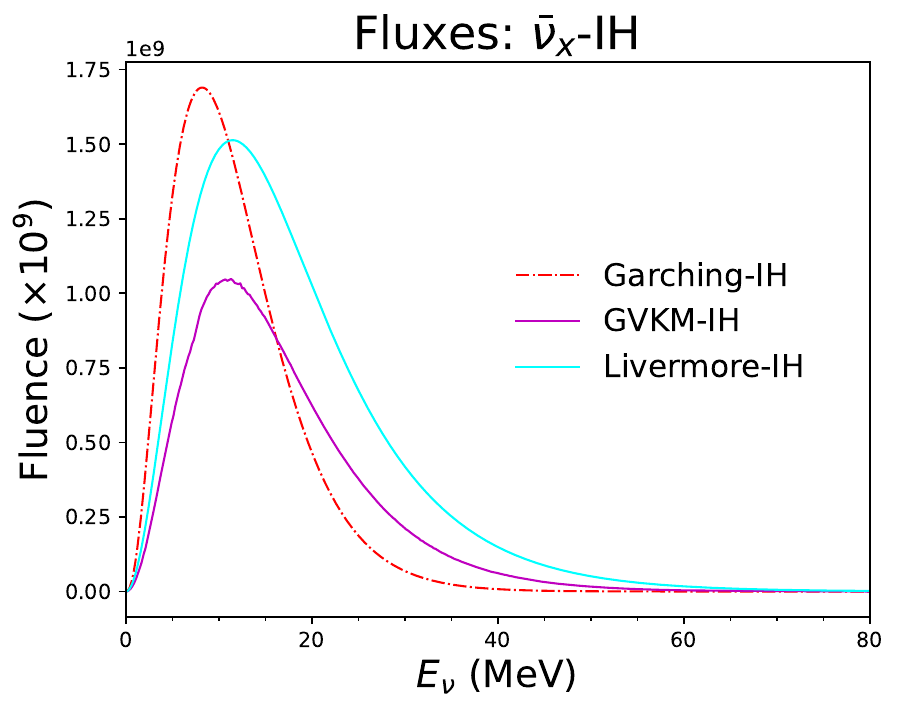}
    \caption{Oscillated fluxes for three different supernova flux models in the inverted mass hierarchy of the neutrinos. In each panel the red curve is for the Garching model, the magenta one is for the GVKM model and the cyan  is for the Livermore model.}
    \label{sn_oscillated_flux_ih}
\end{figure}

In Fig.~\ref{sn_oscillated_flux_nh}, we show the oscillated fluxes corresponding to the three supernova flux models for normal hierarchy of the neutrino masses. For this figure, we have used the fluence equation (\ref{eq-flux})-(\ref{eq-3}) and values of oscillation parameters from Table \ref{table2}. Similar to Fig.~\ref{sn_flux}, the top-left panel is for $\nu_e$, the top-right is for $\bar{\nu}_e$, the bottom-left is for $\nu_x$ and the bottom-right is for $\bar{\nu}_x$. Details of the color coding can be found in the legend.  From the figure, we note that the oscillation of the supernova neutrinos alters the size of the fluences significantly. For $\nu_e$ and $\bar{\nu}_x$, we see that the GVKM and the Livermore models have similar height whereas for $\bar{\nu}_e$ and $\nu_x$, the Garching and the Livermore  models have similar normalization. For the inverted neutrino mass hierarchy in the standard three-flavor scenario, the fluences of $\bar{\nu}_e$ predicted by the GVKM and Livermore models are comparable. In contrast, for $\nu_e$, $\nu_x$, and $\bar{\nu}_x$, these three models predict significantly different fluences, as shown in Fig.~\ref{sn_oscillated_flux_ih}. 

\subsection{Discussion on supernova events}
\label{events}
After the discussion on the fluxes, we next move to the event rates. In Table~\ref{sn_events}, we list the IBD event rates for three different supernova flux models for the supernova distance of 10 kpc. For each model, we present the total number of events for three cases, i.e., unoscillated, normal hierarchy of the neutrino masses and inverted hierarchy of the neutrino masses. From the table we see that the number of expected IBD events varies quite significantly depending on the supernova flux model. The highest number of events can be seen in the case of Livermore model, followed by GVKM model and Garching model. There is also a variation in the number of total events depending upon the true hierarchy of the neutrino masses. Therefore, it is indeed possible to distinguish between the different flux models and to determine the true nature of the neutrino mass hierarchy. For an example of such a study, we refer to one of our previous works \cite{Panda:2023rxa}, where we have studied the expected mass hierarchy sensitivity from a future supernova explosion in Hyper-K and DUNE.

\begin{table}[h]
\centering
\scalebox{1.25}{
\begin{tabular}{|c|c|c|}
\hline
\textbf{Model} & \textbf{Condition} & \textbf{Event rates} \\
\hline
\hline
\textbf{Garching} & Unoscillated & 51,068\\
  \hline
   & NH & 51,742\\
   \hline
    & IH & 53,120 \\
    \hline
    \hline
    \textbf{GVKM} & Unoscillated & 88,528 \\
    \hline
      &  NH & 87,548\\
\hline       
        &  IH  & 85,559 \\
        \hline
        \hline
\textbf{Livermore} & Unoscillated & 148,686 \\
\hline
  &  NH & 163,462\\
  \hline
  & IH & 193,473\\
  \hline
        \hline
\end{tabular}}
\caption{Supernova event rates at 10 kpc distance for three SN models in the standard three-flavor scenario. NH (IH) is normal (inverted) mass hierarchy of the neutrino masses.}
\label{sn_events}
\end{table}

\subsection{Capability to distinguish flux models}

Next we will discuss the capability of the ESSnuSB experiment to distinguish between different flux models. In the previous section, we have shown the expected number of events corresponding to the different supernova flux models. Now we define a Poisson log-likelihood formula to calculate the statistical $\chi^2$,
\begin{equation}
    \chi^2 = 2 \sum_{i=1}^n \left[N_i^{\rm test} - N_i^{\rm true} - N_i^{\rm true} \rm{log} \left( \frac{N_i^{ \rm test}}{N_i^{\rm true}} \right) \right].
    \label{chi}
\end{equation}
In the above formula, we assume event rates of one model as true event rates and $N_i^{\rm test}$ is the event rates of another model with $i$ being the bin index running from 1 to $n$. To incorporate the effect of systematic errors in our calculation, we have taken two types of systematic errors: normalization error and energy calibration error. If we assume a $5\%$ systematic uncertainty in both normalization and energy calibration, then the modified expression for $N_{i}^{\rm test}$ becomes

 \begin{equation}
    N_i^{\rm test} \rightarrow N_i^{\rm test} [( 1+ 0.05 \zeta_1 ) + 0.05 \zeta_2(E_i^{\prime} - \bar{E^{\prime}})/(E^{\prime}_{\rm max}-E^{\prime}_{\rm min})],
    \label{error}
\end{equation}
where $\zeta_1$ and $\zeta_2$ are the pull variables associated with normalization and energy calibration uncertainties, respectively. Here, $E^{\prime}_{\rm max}$ and $E^{\prime}_{\rm min}$ denote the maximum and minimum energies of the event spectrum and $\bar{E}^{\prime} = \frac{1}{2}(E^{\prime}_{\rm max} + E^{\prime}_{\rm min})$ represents the midpoint of the energy interval, and $E_i^{\prime}$ is the reconstructed energy of the $i$th bin.

Thus, the total sensitivity (statistical + systematic) can be expressed as,
\begin{equation}
    \chi^2_{ \rm total} = \chi^2 + \zeta_1^2 +\zeta_2^2 \;.
    \label{chi-sys}
\end{equation}

\begin{figure}[h]
    \centering
    \includegraphics[width=42mm, height=45mm]{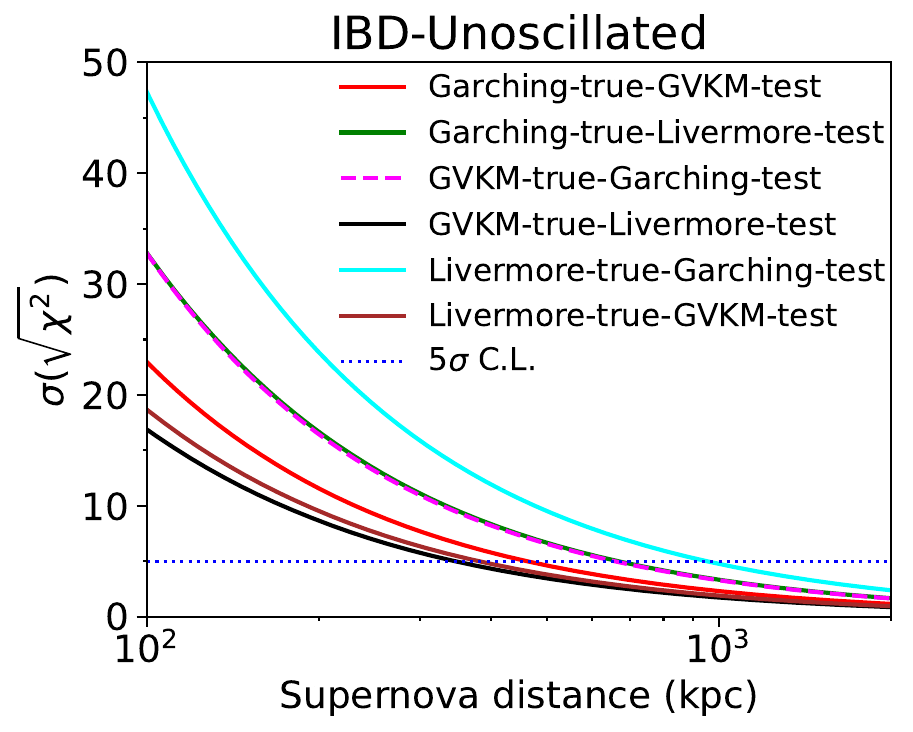}
    \includegraphics[width=42mm, height=45mm]{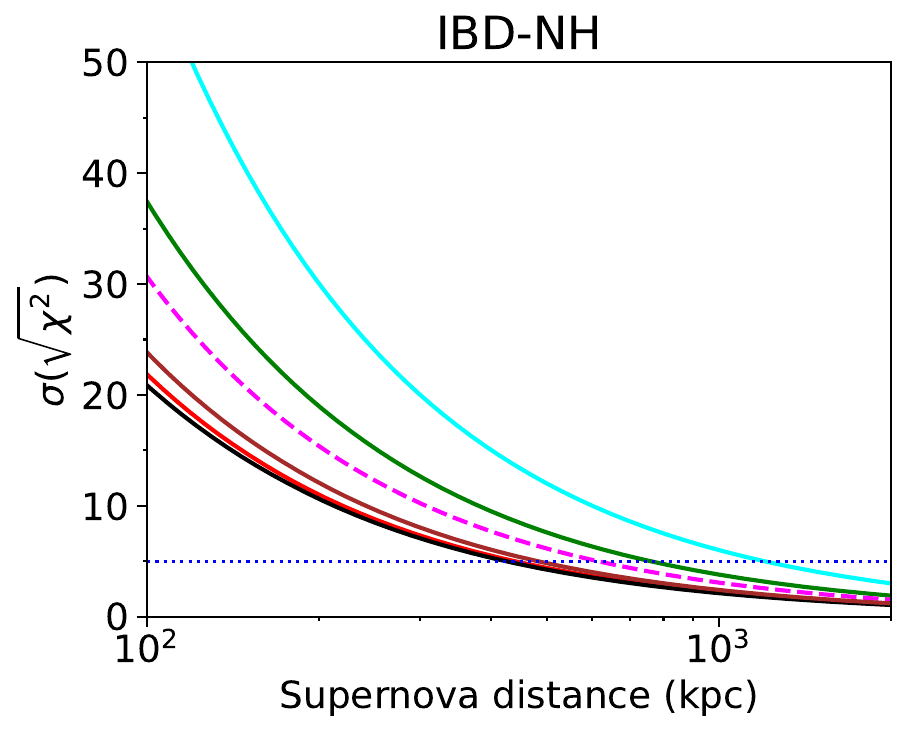}
    \includegraphics[width=42mm, height=45mm]{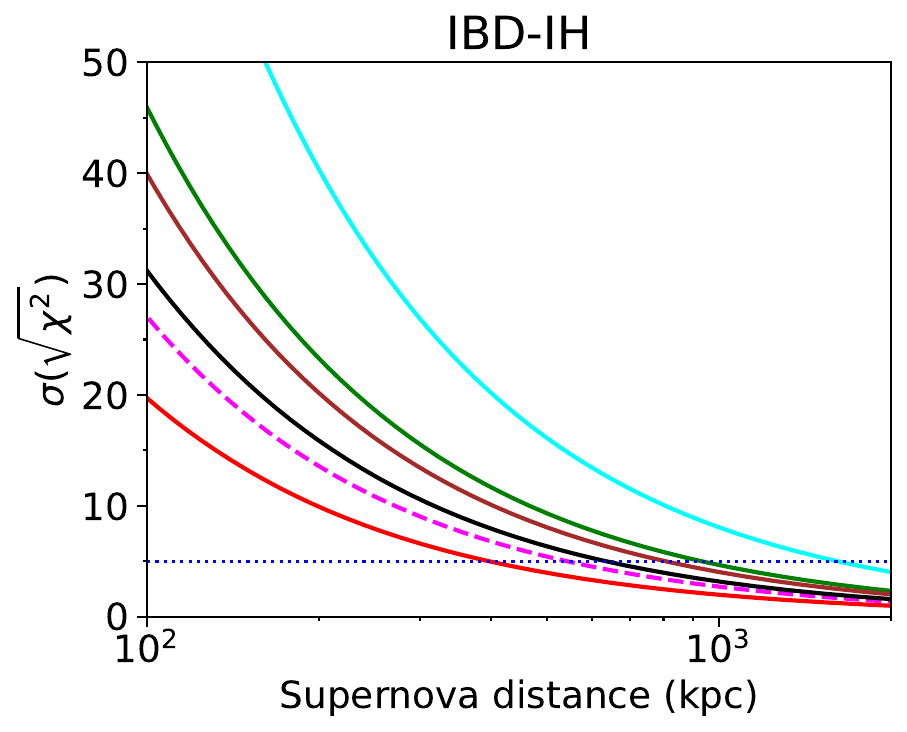}
    \caption{Sensitivity of ESSnuSB to distinguish different supernova flux models as a function of supernova distance. In each panel, the different curves represent the sensitivity to distinguish between pairs of models. The legends in each panel specify the true and test spectra for the corresponding curves. For instance, the red curve in each panel represents the sensitivity to distinguish the Garching model from the GVKM model, with Garching model taken in the true spectrum and GVKM model in the test spectrum.}
    \label{sn_sens1}
\end{figure}

In Fig.~\ref{sn_sens1}, we show the sensitivity of the ESSnuSB FD to distinguish different supernova flux models as a function of supernova distance in kpc. The left panel is without assuming any oscillations, the middle panel is for normal hierarchy of the neutrino masses and the right panel is for the inverted hierarchy of the neutrino masses. In this figure, we assume the detector efficiency to be 100\%, energy resolution to be 10\% and systematic errors to be 5\% for both normalization and energy calibration. From the figure, we can see that one can differentiate different combinations of fluxes from each other at $5\sigma$ C.L. if a supernova explosion occurs at a distance between 300 kpc to 1000 kpc from Earth. It can also be observed that the cyan curve shows a very high sensitivity implying the fact that the Livermore model can be distinguished from the Garching model with a very high confidence level. As this curve is clearly separated from the other curves, in the subsequent figures, we will omit this curve. 

\begin{figure}[h]
    \centering
    \includegraphics[width=42mm, height=45mm]{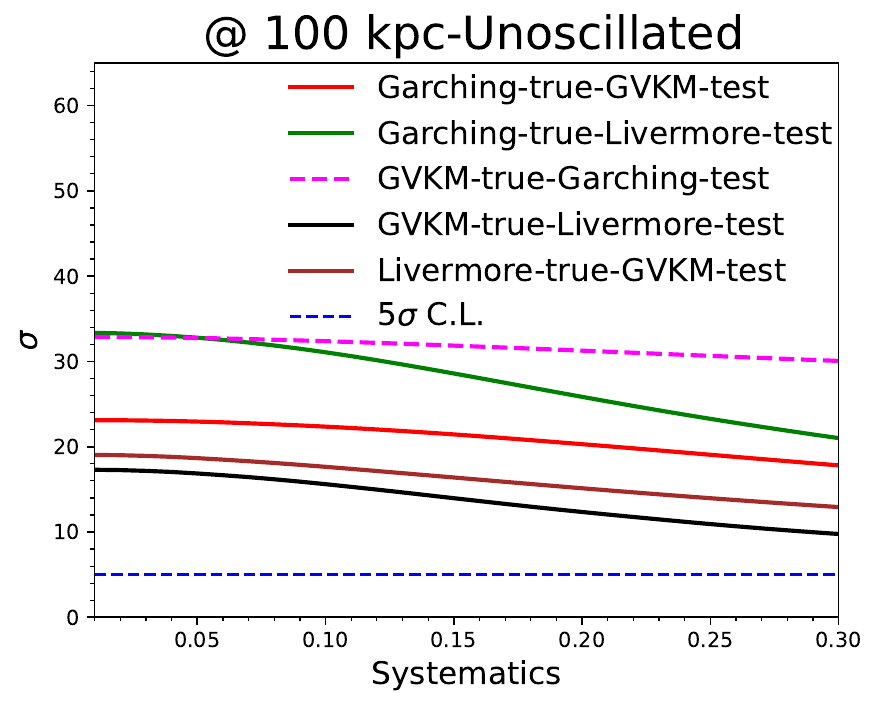}
    \includegraphics[width=42mm, height=45mm]{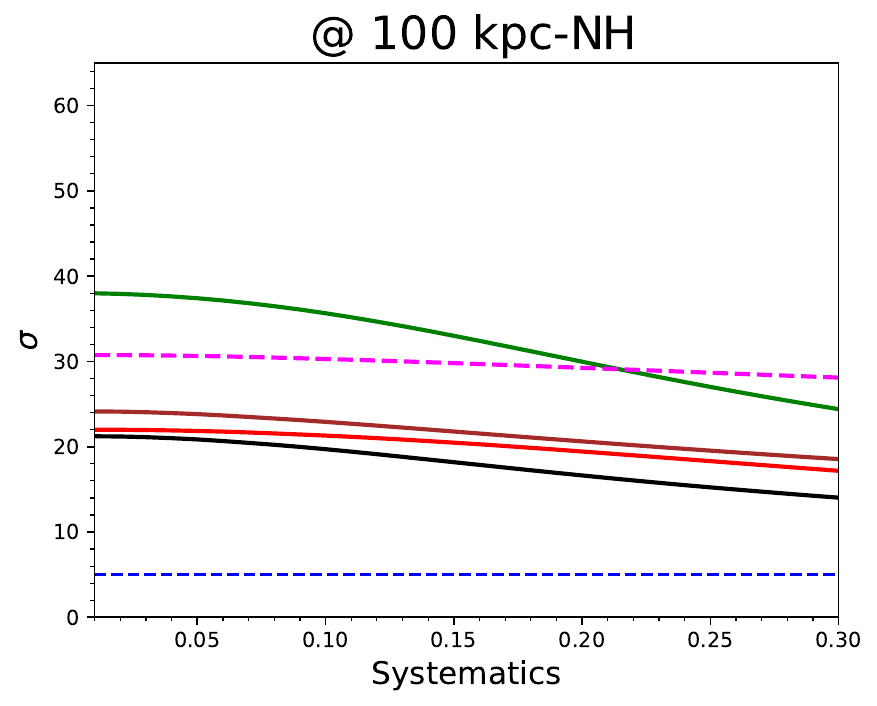}
    \includegraphics[width=42mm, height=45mm]{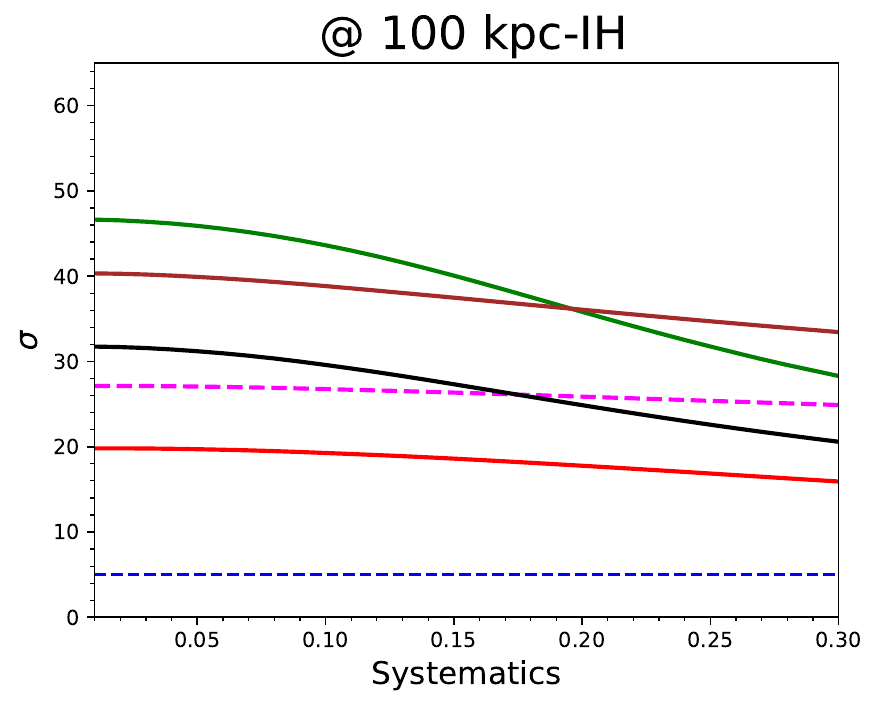}
    \caption{Sensitivity of ESSnuSB to distinguish different supernova flux models as a function of systematic error.}
    \label{sn_sens2}
\end{figure}

In Fig.~\ref{sn_sens2}, we show the sensitivity of the ESSnuSB to distinguish different supernova flux models as a function of systematic errors. In this case, both normalization and energy calibration errors are varied. Similar to Fig.~\ref{sn_sens1}, the left panel is without assuming any oscillations, the middle panel is for normal hierarchy of the neutrino masses and the right panel is for the inverted hierarchy of the neutrino masses. In each panel, different curves demonstrate sensitivity to distinguish one model from another. In this figure, we assume the detector efficiency to be 100\%, energy resolution to be 10\% and distance of the supernova to be 100 kpc. From the figure, we can see that as the systematic error increases, the sensitivity decreases. 

\begin{figure}[h]
    \centering
    \includegraphics[width=42mm, height=45mm]{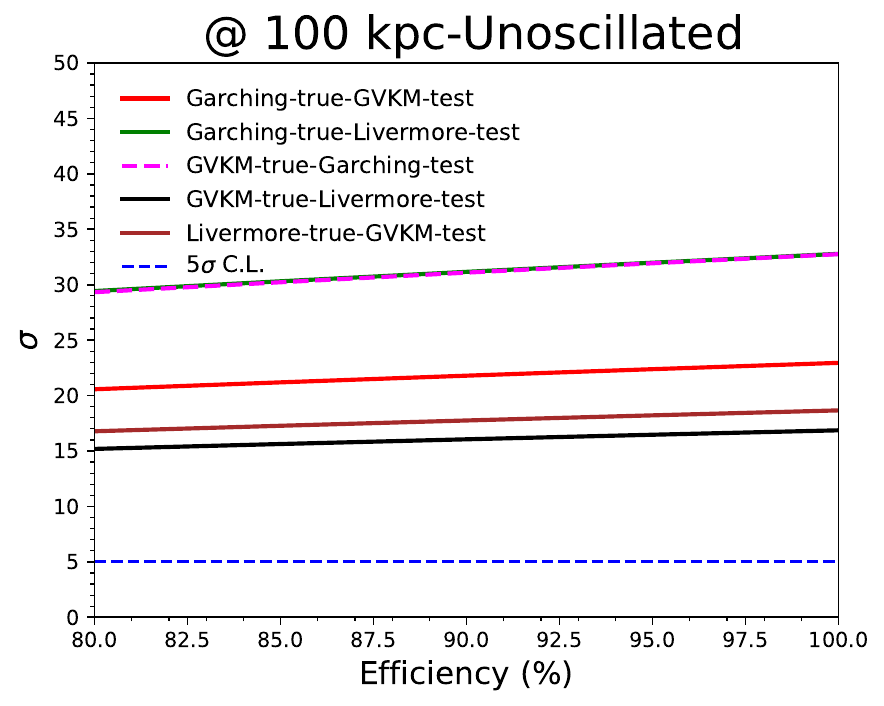}
    \includegraphics[width=42mm, height=45mm]{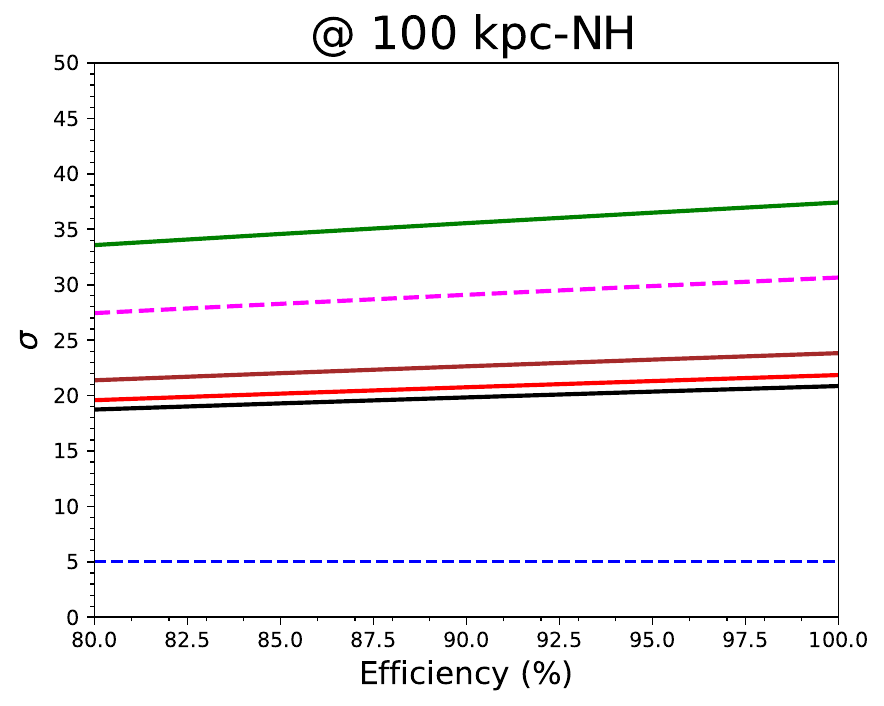}
    \includegraphics[width=42mm, height=45mm]{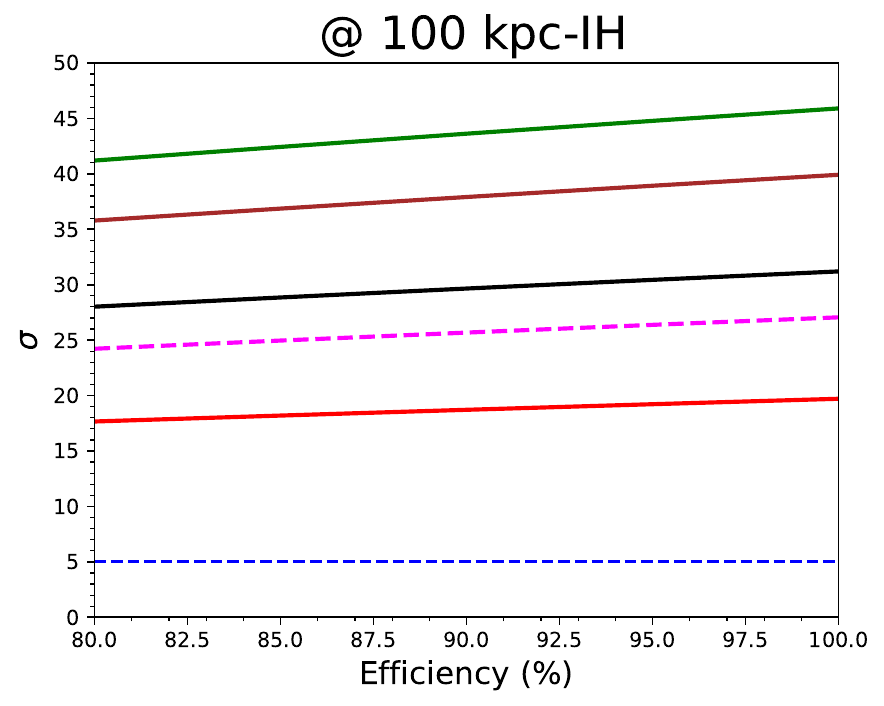}
    \caption{Sensitivity of ESSnuSB to distinguish different supernova flux models as a function of efficiency.}
    \label{sn_sens3}
\end{figure}

In Fig.~\ref{sn_sens3}, we show the sensitivity of the ESSnuSB to distinguish different supernova flux models as a function of detector efficiency. The left panel is without assuming any oscillations, the middle panel is for normal hierarchy of the neutrino masses and the right panel is for the inverted hierarchy of the neutrino masses.
Different curves demonstrate sensitivity to distinguish one model from another. In this figure, we assume the energy resolution to be 10\%, distance of the supernova to be 100 kpc and systematic errors to be 5\% for both normalization and energy calibration. From this panel we can see that the sensitivity improves as the detector efficiency increases.  

\begin{figure}[h]
    \centering
    \includegraphics[width=42mm, height=45mm]{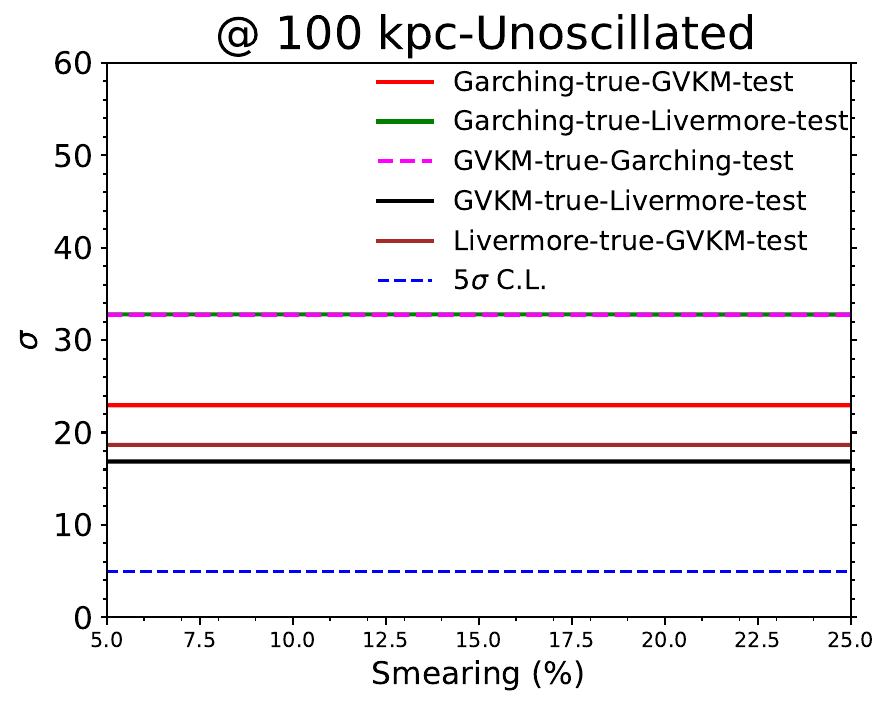}
    \includegraphics[width=42mm, height=45mm]{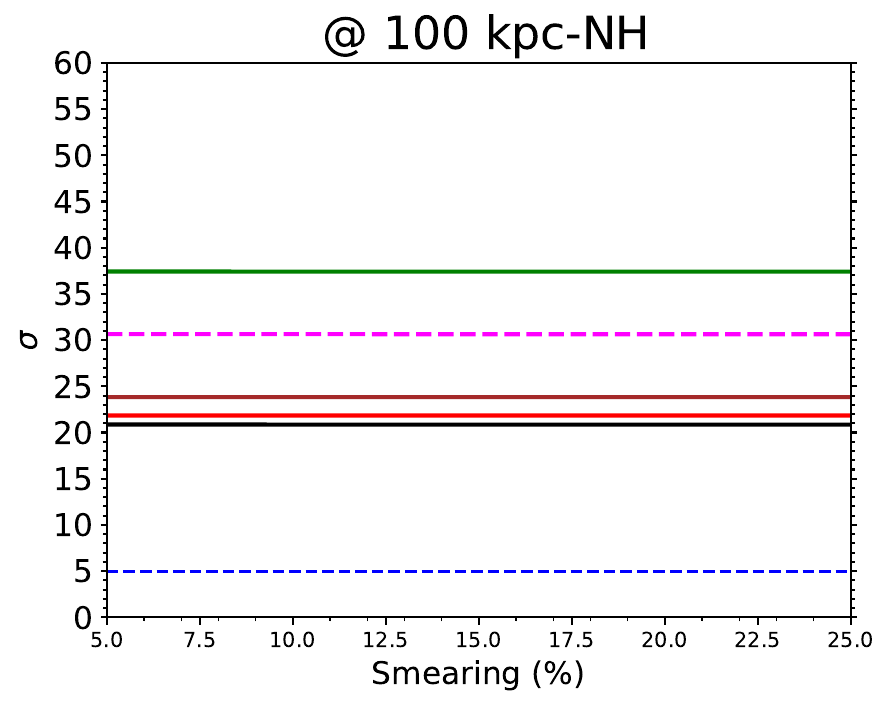}
    \includegraphics[width=42mm, height=45mm]{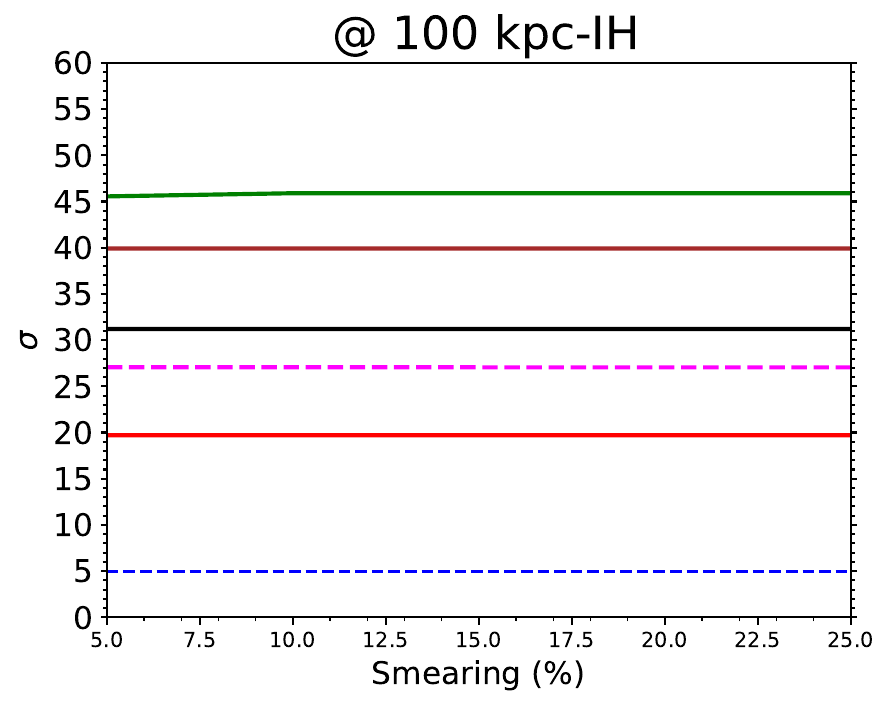}
    \caption{Sensitivity of ESSnuSB to distinguish different supernova flux models as a function of smearing.}
    \label{sn_sens4}
\end{figure}

In Fig.~\ref{sn_sens4}, we show the sensitivity of the ESSnuSB to distinguish different supernova flux models as a function of energy resolution. The left panel is without assuming any oscillations, the middle panel is for normal hierarchy of the neutrino masses and the right panel is for the inverted hierarchy of the neutrino masses. Different curves demonstrate sensitivity to distinguish one model from another. In this figure, we assume the detector efficiency to be 100\%, distance of the supernova to be 100 kpc and systematic errors to be 5\% for both normalization and energy calibration. From this panel we can see that the sensitivity remains almost flat as the energy resolution changes.

\section{Summary and future scope}

In this paper, we have estimated the sensitivity of the ESSnuSB FD to supernova neutrinos in terms of different flux models. The flux models that we considered in our analysis are: Livermore, GVKM and Garching. First we have shown that although the shape of the fluxes from different models are similar, the normalization is very different. For $\nu_e$ and $\bar{\nu}_e$, all the three models have different size of the fluxes, whereas for $\nu_x$ and $\bar{\nu}_x$, the predictions of the GVKM and Livermore are similar but the prediction of the Garching model is very different. Next we have shown that the oscillations of supernova neutrinos alter the size of the fluences significantly. We have further shown that the number of expected IBD events varies quite significantly depending on the supernova flux model. The highest number of events can be seen in the case of Livermore model, followed by GVKM model and Garching model. There is also variation in the number of total events depending upon the true hierarchy of the neutrino masses. Our results show that if a supernova explosion occurs at a distance between 300 kpc to 1000 kpc from Earth, one can differentiate different combinations of fluxes from each other at $5 \sigma$ C.L. Among the three models, the Livermore model can be distinguished from the Garching model with a very high confidence level.  The sensitivity generally decreases with large systematic errors while it improves with better detection efficiency. Finally, we have also shown the impact of energy resolution on the sensitivity and found that the effect is minimal when the energy resolution is varied from $5\%$ to $25\%$.

In future, there is a scope to perform a detailed detector simulation optimized for the supernova neutrinos in order to estimate the realistic detector performance in terms of detector efficiency and energy resolution. In this context it would be very interesting to see the effect of Gd doping. 
Gd doping can improve the detection efficiency and background rejection of the supernova neutrinos by enabling efficient neutron tagging. Though an qualitative effect could be understood from our Fig.~\ref{sn_sens3}, where we have shown effect of efficiencies in distinguishing different flux models, a detailed simulation including Gd will be important to estimate more realistic sensitivities. The study of supernova neutrinos is a broad field, and in the future, the ESSnuSB far detector can be used for various physics studies, such as exploring collective effects, performing precision measurements of standard oscillation parameters, and investigating new physics scenarios for example non-standard interactions, sterile neutrinos etc. In our present analysis, we have considered only three supernova flux models but there exists many state-of-the-art flux models \cite{github1} which one can study at ESSnuSB. 

\section*{Funding}

Funded by the European Union, Project 101094628. Views and opinions expressed are however those of the author(s) only and do not necessarily reflect those of the European Union. Neither the European Union nor the granting authority can be held responsible for them. We acknowledge further support provided by the following research funding agencies: Centre National de la Recherche Scientifique, France; Deutsche Forschungsgemeinschaft
Projektnummer 423761110, and under the Excellence Strategy of the Federal Government and the Länder, Germany; Ministry of Science and Education of Republic of Croatia grant No. PK.1.1.10.0002; the European Union’s Horizon 2020 research and innovation programme under the Marie Skłodowska-Curie grant agreement No 860881-HIDDeN; the European Union NextGenerationEU, through the National Recovery and Resilience Plan of the Republic of Bulgaria, project No. BG-RRP-2.004-0008-C01; Roland Gustafssons Stiftelse för teoretisk fysik, Sweden; Swiss National Science Foundation (SNSF), Croatian Science Foundation (HRZZ) and National Recovery and Resilience Plan (NPOO) via the grants MAPS IZ11Z0$\_$230193 and DOK-NPOO-2023-10-1262; as well as support provided by the universities and laboratories to which the authors of this report are affiliated, see the author list on the first page.

\vspace{6pt}

\authorcontributions{M. Ghosh, R. Mohanta and P.Panda are the corresponding authors. They have contributed equally in conceptualization and writing. Results are generated by P. Panda. Other authors have contributed in reviewing the results and the manuscript.}

\dataavailability{The data are available from the
authors upon reasonable request.} 

\conflictsofinterest{We do not have any conflict of interest.} 
\appendixtitles{no} %
\appendixstart
\appendix
\section[\appendixname~\thesection]{Authorship of the ESSnuSB Collaboration}

J.~Aguilar $^{1}$, M.~Anastasopoulos $^{2}$, D.~Barčot $^{3}$, E.~Baussan$^{4}$, A.K.~Bhattacharyya $^{2}$, A.~Bignami $^{2}$, M.~Blennow $^{5,6}$, M.~Bogomilov $^{7}$, B.~Bolling $^{2}$, E.~Bouquerel $^{4}$, F.~Bramati $^{8}$, A.~Branca $^{8}$, G.~Brunetti $^{8}$, I.~Bustinduy $^{1}$, C.J.~Carlile $^{9}$, J.~Cederkall $^{9}$, T.~W.~Choi $^{10}$, S.~Choubey $^{5,6}$, P.~Christiansen $^{9}$, E.~Cristaldo Morales $^{8}$, P.~Cupia\l $^{12}$, D.~D'Ago $^{13}$, H.~Danared $^{2}$, J.~P.~A.~M.~de~Andr\'{e} $^{4}$, M.~Dracos $^{4}$, I.~Efthymiopoulos $^{14}$, T.~Ekel\"{o}f $^{10}$, M.~Eshraqi $^{2}$, G.~Fanourakis $^{15}$, A.~Farricker $^{16}$, E.~Fasoula $^{17,18}$, T.~Fukuda $^{19}$, S.~Gago $^{1}$, N.~Gazis $^{2}$, Th.~Geralis $^{15}$, M.~Ghosh $^{3}$*, A.~Giarnetti $^{20}$, G.~Gokbulut $^{21,22}$, C.~Hagner $^{23}$, L.~Halić $^{3}$, J.~Hiegel $^{4}$, M.~Hooft $^{22}$, K.~E.~Iversen $^{9}$, N.~Jachowicz $^{22}$, M.~Jakkapu $^{3}$, M.~Jensen $^{2}$, I.~Karakoulias $^{15,17}$, E.~Kasimi, $^{17}$, A.~Kayis Topaksu $^{21}$, B.~Kliček $^{3}$, K.~Kordas $^{17,18}$, B.~Kovač $^{3}$, A.~Leisos $^{24}$, A.~Longhin $^{25}$, M.~López $^{1}$, C.~Maiano $^{2}$, S.~Marangoni $^{8}$, J.~G.~Marcos $^{22}$, C.~Marrelli $^{14}$, D.~Meloni $^{20}$, M.~Mezzetto $^{13}$, N.~Milas $^{2}$, R.~Mohanta $^{26}$*, J.L.~Muñoz $^{1}$, K.~Niewczas $^{22}$, M.~Oglakci $^{21}$, T.~Ohlsson $^{5,6}$, M.~Olveg\r{a}rd $^{10}$, P.~Panda $^{26}$*, M.~Pari $^{24}$, D.~Patrzalek $^{2}$, G.~Petkov $^{7}$, Ch.~Petridou $^{17,18}$, P.~Poussot $^{4}$, A.~Psallidas $^{15}$, F.~Pupilli $^{13}$, D.~Saiang $^{27}$, D.~Sampsonidis $^{17,18}$, A.~Scanu $^{8}$, C.~Schwab $^{4}$, F.~Sordo $^{1}$, G.~Stavropoulos $^{15}$, M.~Stipčević $^{3}$, R.~Tarkeshian $^{2}$, F.~Terranova $^{8}$, T.~Tolba $^{23}$, M.~Topp-Mugglestone $^{14}$, E.~Trachanas $^{2}$, R.~Tsenov $^{7}$, A.~Tsirigotis $^{24}$, S.~E.~Tzamarias $^{17}$, M.~Vanderpoorten $^{22}$, G.~Vankova-Kirilova $^{7}$, N.~Vassilopoulos $^{28}$, S.~Vihonen $^{5,6}$, J.~Wurtz $^{4}$, V.~Zeter $^{4}$ and O.~Zormpa $^{15}$

\subsection*{Affiliations}
\begin{itemize}
\item[$^{1}$] Consorcio ESS-bilbao, Parque Científico y Tecnológico de Bizkaia, Laida Bidea, Edificio 207-B, 48160 Derio, Bizkaia, Spain
\item[$^{2}$] European Spallation Source, Box 176, SE-221 00 Lund, Sweden 
\item[$^{3}$] Center of Excellence for Advanced Materials and Sensing Devices, Ruđer Bo\v{s}kovi\'c Institute, 10000 Zagreb, Croatia 
\item[$^{4}$] IPHC, Universit\'{e} de Strasbourg, CNRS/IN2P3, Strasbourg, France 
\item[$^{5}$] Department of Physics, School of Engineering Sciences, KTH Royal Institute of Technology,Roslagstullsbacken 21, 106 91 Stockholm, Sweden 
\item[$^{6}$] The Oskar Klein Centre, AlbaNova University Center, Roslagstullsbacken 21, 106 91 Stockholm, Sweden 
\item[$^{7}$] Sofia University St. Kliment Ohridski, Faculty of Physics, 1164 Sofia, Bulgaria 
\item[$^{8}$] University of Milano-Bicocca and INFN Sez. di Milano-Bicocca, 20126 Milano, Italy 
\item[$^{9}$] Department of Physics, Lund University, P.O Box 118, 221 00 Lund, Sweden 
\item[$^{10}$] Department of Physics and Astronomy, FREIA Division, Uppsala University, P.O. Box 516, 751 20 Uppsala, Sweden 
\item[$^{11}$] Faculty of Engineering, Lund University, P.O Box 118, 221 00 Lund, Sweden 
\item[$^{12}$] AGH University of Krakow, al. A. Mickiewicza 30, 30-059 Krakow, Poland 
\item[$^{13}$] INFN Sez. di Padova, Padova, Italy 
\item[$^{14}$] CERN, 1211 Geneva 23, Switzerland 
\item[$^{15}$] Institute of Nuclear and Particle Physics, NCSR Demokritos, Neapoleos 27, 15341 Agia Paraskevi, Greece 
\item[$^{16}$] Cockroft Institute (A36), Liverpool University, Warrington WA4 4AD, UK 
\item[$^{17}$] Department of Physics, Aristotle University of Thessaloniki, Thessaloniki, Greece 
\item[$^{18}$] Center for Interdisciplinary Research and Innovation (CIRI-AUTH), Thessaloniki, Greece 
\item[$^{19}$] Institute for Advanced Research, Nagoya University, Nagoya 464–8601, Japan 
\item[$^{20}$] Dipartimento di Matematica e Fisica, Universit\'a di Roma Tre, Via della Vasca Navale 84, 00146 Rome, Italy 
\item[$^{21}$] University of Cukurova, Faculty of Science and Letters, Department of Physics, 01330 Adana, Turkey 
\item[$^{22}$] Department of Physics and Astronomy, Ghent University, Proeftuinstraat 86, B-9000 Ghent, Belgium 
\item[$^{23}$] Institute for Experimental Physics, Hamburg University, 22761 Hamburg, Germany 
\item[$^{24}$] Physics Laboratory, School of Science and Technology, Hellenic Open University, 26335, Patras, Greece 
\item[$^{25}$] Department of Physics and Astronomy "G. Galilei", University of Padova and INFN Sezione di Padova, Italy 
\item[$^{26}$] School of Physics, University of Hyderabad, Hyderabad - 500046, India 
\item[$^{27}$] Department of Civil, Environmental and Natural Resources Engineering, Luleå University of Technology, SE-971 87 Lulea, Sweden 
\item[$^{28}$] Institute of High Energy Physics (IHEP) Dongguan Campus, Chinese Academy of Sciences (CAS), Guangdong 523803, China
\end{itemize}

\begin{adjustwidth}{-\extralength}{0cm}

\reftitle{References}

 \bibliography{sn-bibliography}

\PublishersNote{}

\end{adjustwidth}
\end{document}